\documentclass[a4paper,10pt,twocolumn]{article} 
\usepackage{graphicx}
\usepackage{subfigure}
\usepackage{amsmath}
\usepackage{amssymb}
\usepackage{wrapfig}
\usepackage{color}
\usepackage[english]{babel}
\usepackage[warning,math]{easyeqn}
\usepackage[thinlines]{easybmat}
\usepackage{enumerate} 
\usepackage{multirow} 
\usepackage[pdftex,
            pdfauthor={Yves-Henri Sanejouand},
            pdftitle={Xenon binding sites in the NMDA receptor},
            pdfsubject={NMDAR},
            pdfkeywords={Anesthesia,NMDA receptor,Xenon,Molecular Dynamics,Absolute Binding Free Energy},
            pdfproducer={Latex with hyperref},
            pdfcreator={Texmaker}]{hyperref}
\begin{document}

\title{\textbf{At least three xenon binding sites \\
in the glycine binding domain of the \\
N-methyl D-aspartate receptor}}

\author{Yves-Henri Sanejouand\footnote{yves-henri.sanejouand@univ-nantes.fr}\\ \\
		US2B, UMR 6286 of CNRS,\\
        Nantes University, France.} 
\date{March 4$^{th}$, 2022}
\maketitle

\section*{Abstract}

Xenon can produce general anesthesia.
Its main protein target is the N-methyl-D-aspartate receptor, a ionotropic channel playing a pivotal role in the function of the central nervous system. 

The molecular mechanisms allowing this noble gas to have such
a specific effect remain obscure, probably as a consequence of the lack of structural data at the atomic
level of detail.
Herein, as a result of five independent molecular dynamics simulations, three different binding sites were found for xenon in the glycine binding domain of the 
N-methyl-D-aspartate receptor.
The absolute binding free energy of xenon in these sites
ranges between -8 and -14 kJ$\cdot$mole$^{-1}$. However,
it depends significantly upon the protein conformer chosen
for performing the calculation, suggesting that larger values
could probably be obtained, if other conformers were considered.

These three sites are next to each other,
one of them being next to the glycine site.
This could explain why the F758W and F758Y mutations can prevent competitive inhibition by xenon
without affecting glycine binding.

\vskip 1cm
\textbf{Keywords}: Xenon, NMDA receptor, Molecular Dynamics, Absolute Binding Free Energy.

\section*{Introduction}

Xenon can produce general anesthesia without causing undesirable side effects \cite{Erdmann:90,Morita:03}. 
This is likely due to the fact that
xenon potently inhibits the excitatory N-methyl-D-aspartate receptor (NMDAR) \cite{Lieb:98}, a ionotropic channel which is a coincidence detector able to detect synchronicity 
in synaptic depolarization events \cite{Sjostrom:06,Kohr:06}. 
Interestingly, xenon has other attractive pharmacological properties \cite{Maze:03,Franks:10}. For instance, it provides neuroprotection against hypoxia--ischemia \cite{Dickinson:10}.

Though the main protein target of xenon has been known for twenty-five years \cite{Lieb:98},
the molecular mechanisms allowing this noble gas to have such specific effects remain obscure, probably as a consequence of the lack of structural data at the atomic level of detail. Indeed, though numerous crystal structures of proteins with bound xenon atoms have been obtained \cite{Petsko:84,Fourme:98,Abraini:07,Colloch:16}, in the case of the NMDAR, no such data is presently available in the Protein Data Bank \cite{RCSB}.    

Thanks to computational approaches, the location of the xenon binding site(s) in the NMDAR has however been looked for a couple of times. Using a Grand Canonical Monte Carlo method, it was first proposed to be within the glycine binding site itself \cite{Franks:07}. However, 
two mutations in this site \cite{1PBQ,Zefirov:03}, namely, F758W and F758Y, were next found to prevent
competitive inhibition by xenon without affecting glycine binding \cite{Dickinson:12}.
In a subsequent study of the ligand binding domains of NMDAR, performed using docking, molecular dynamics (MD) simulations and absolute binding free energy (ABFE) calculations, four putative xenon binding sites were identified, one nearby the glutamate site, the three others far from the agonist binding sites \cite{Tang:10}, that is, all far from F758. 

Though the effects of the F758W and F758Y mutations could prove to have an allosteric character \cite{Jane:12,Paoletti:13}, that is, to be long-range ones, it is also possible that both previous studies have missed the actual binding site(s) of xenon in the NMDAR.   
For instance, both studies were performed starting from
crystal structures of holo forms (PDB 1PBQ \cite{1PBQ} or 2A5T \cite{2A5T}), while xenon could for instance inhibit glycine binding by stabilizing an apo form. 

Also, in the later study, putative xenon binding sites were first located using standard docking methods \cite{Tang:10}. But since xenon is an apolar atom, it can bind to any large enough hydrophobic cavity, whose size and shape can in turn fluctuate, as a consequence
of protein intrinsic flexibility \cite{Mouawad:06,Okuno:18,Blondel:19}, which is usually little accounted for in most docking protocols \cite{Abagyan:08,Wolfson:10}.

On the other hand, it has been shown that it is nowadays possible to dock accurately small ligands on proteins using explicit solvent MD simulations \cite{Shaw:11c,Zacharias:14,Lau:18,Caflisch:18,Mondal:21}, that is, with a much more detailled and accurate energy function, without making any assumption about the location of the binding site or about the rigidity of the protein backbone. 

So, the main goal of the present work was to locate the binding site(s) of xenon in the NMDAR, using explicit solvent MD simulations only. Then, in order to obtain a lower bound for the affinity of xenon for the putative binding sites thus found, state-of-the-art ABFE calculations were performed. 

The NMDAR is an obligatory heterotetramer usually made of 
two GluN1 and two GluN2 subunits, glycine binding on the GluN1 subunits while glutamate binds on the GluN2 ones \cite{Farrant:01}. Since xenon is expected to bind in the vicinity of 
the glycine binding site \cite{Franks:07,Dickinson:12},
only the glycine binding domain of the NMDAR, for which a high resolution crystal structure of an apo form is available \cite{4kcc}, is considered hereafter.

\section*{Methods}

\subsection*{Multiple sequence alignment}

Starting from the sequence of the glycine binding domain of the GluN1 subunit of the rat NMDAR (gene GRIN1), as found in PDB 4KCC \cite{4kcc},
977 sequences were retrieved from the Uniprot database \cite{Uniprot}, and a multiple alignment was performed, using Clustal$\Omega$ \cite{Clustalo}. Then, 648 sequences more than 66\% identical (87 $\pm$ 7\%, on average) to the complete (without any gap --see below) rat sequence of the glycine binding domain were retained for further analysis. These sequences come from 237 species such as human, chicken or the african clawed frog. 

\subsection*{Molecular dynamics simulations}

\label{sec:md}
Molecular dynamics simulations were performed with gromacs \cite{Gromacs} version 2018.6, using the Amber 99SB-ILDN forcefield \cite{Shaw:10amb} and the TIP3P water model \cite{Jorgensen:83}. 
Short range electrostatic and van der Waals interactions were cut off at 1.2 nm, long-range electrostatics being handled through the particle mesh Ewald method \cite{Darden:93}. The xenon Lennard-Jones interaction parameters are the following ones: 
$\sigma_{Xe}$ = 0.4063 nm, $\epsilon_{Xe}$ = 2.35 kJ $\cdot$ mole$^{-1}$ \cite{Masters:15,Vrabec:19}.

Missing loops 441-448 and 491-495 were added to the apo conformation of the glycine binding domain of the rat NMDAR (PDB 4KCC \cite{4kcc}, R=1.89{\AA}), using the REMODEL protocol of ROSETTA \cite{Rosetta:15}. Then, ten xenon atoms were randomly added around the protein, so as to be at least 0.5 nm away from any heavy protein or other xenon atom. 
Next, together with the 226 
water molecules found in the crystal structure, the protein was embedded in a water box, with its boundaries at least 1 nm away from the protein, the box volume being of 707 nm$^3$. Finally, sodium and chloride ions were added so as to neutralize the charge of the system and to reach a salt concentration of 150 mM$\cdot$L$^{-1}$, 
as often done \cite{Okuno:18,Schulten:09}.
The system was then relaxed using steepest descent minimization, with harmonic restraints (force constant of 1000 kJ$\cdot$mol$^{-1}\cdot$nm$^{-2}$) on protein heavy atoms, until a maximum force threshold of 1000 kJ$\cdot$mol$^{-1}\cdot$nm$^{-1}$ was reached. 
The solvent was next equilibrated at 300$^\circ$K, first during one nsec with a control of both volume and temperature, using the Berendsen thermostat \cite{Berendsen:84} and a coupling constant of 0.1 ps, then during another nsec with a control of both pressure and temperature, using the Parrinello-Rahman barostat \cite{Parrinello:81} and a coupling constant of 2 ps. 
Finally, restraints were removed and 500 ns of simulation at 300$^\circ$K were
performed, with a timestep of 2 fs and all bonds constrained using the LINCS
algorithm \cite{LINCS}, 1000 frames being saved for further analysis.

Five MD simulations (coined A to D) are considered hereafter, each of them starting with a different configuration of the ten xenon atoms around the NMDAR. 

\begin{figure}[t!]
\hskip 0.3 cm
\includegraphics[width=7.0 cm]{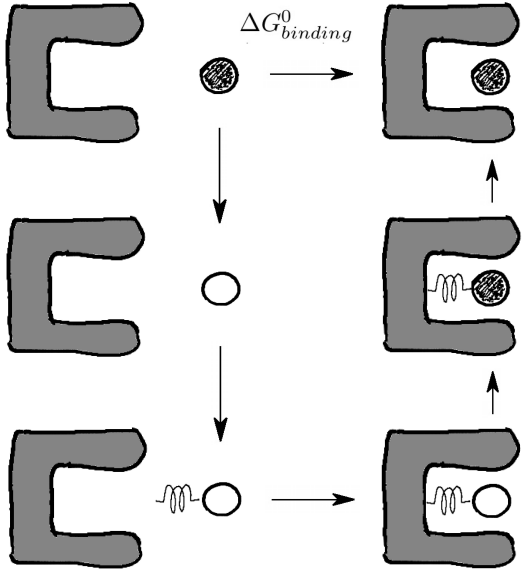}
\caption[]{
Thermodynamical cycle used to compute the absolute binding free energy of xenon.
Top left: the xenon atom (filled circle) is first annihilated (empty circle). An harmonic spring is then added in order
to restrain its motion. Bottom: the harmonic spring is linked to a protein atom.
Right: the xenon atom is restored. Then, the harmonic spring is removed.
}
\label{Fig:dgcycle}
\end{figure}

\subsection*{Absolute binding free energy}

$\Delta G^0_{binding}$, the absolute binding free energy (ABFE) of xenon in the NMDAR, was calculated through a thermodynamical cycle (see Fig. \ref{Fig:dgcycle}), that is, instead of obtaining it directly, the calculation was performed in several steps, so that \cite{Shankar:86,Smith:96}:
\begin{multline}
$$
\Delta G^0_{binding} = \Delta G^{solv}_{annihil} + \Delta G^{solv}_{restr} + \\
\Delta G^0_{transf} + \\
\Delta G^{prot}_{creation} + \Delta G^{prot}_{unrestr} \hspace{1.4cm}
\notag
$$
\end{multline}
where $\Delta G^{solv}_{annihil}$ is the free energy of transfer of the xenon atom from the bulk to the gas, that is, the opposite of its solvation free energy, $\Delta G^{solv}_{restr}$, the free energy cost for adding an harmonic restraint, $\Delta G^0_{transf}$, the free energy of transfer of the restrained, annihilated, xenon atom from the bulk to the protein, where it is linked to a given protein atom, $\Delta G^{prot}_{creation}$ being the free energy of transfer of the restrained xenon atom from the gas to the protein, while $\Delta G^{prot}_{unrestr}$ is the free energy cost for removing the harmonic restraint.

$\Delta G^{solv}_{annihil}$ can be obtained from experimental data, namely: -5.4 kJ$\cdot$mole$^{-1}$ \cite{Marcus:84,Nau:18}.  
On the other hand:
\[
\Delta G^0_{transf} = \textnormal{0}
\]
while \cite{Smith:96b,McCammon:97,Ren:17}:
\[
\Delta G^{solv}_{restr} = -RT \ln \left[ C_0 \left( \frac{2 \pi RT}{k_{restr}} \right)^{\frac{3}{2}} \right] 
\]
where $k_{restr}$ is the force constant of the harmonic restraint, $T$, the temperature, $R$, the molar gas constant, $C_0$ being the standard state concentration (1 mol$\cdot$L$^{-1}$).  
Herein, $k_{restr}=$ 418 kJ$\cdot$mole$^{-1}\cdot$nm$^{-2}$, T=300$^\circ$K, so: 
$\Delta G^{solv}_{restr} = \textnormal{7.8 kJ}\cdot\textnormal{mole}^{-1}$.

The last two terms,
$\Delta G^{prot}_{creation}$ and $\Delta G^{prot}_{unrestr}$, were calculated, as well as their estimated statistical uncertainty, using the Bennett acceptance ratio method \cite{Bennett:76,Boresch:11,Mobley:15}, as implemented in the g\_bar tool of gromacs \cite{Gromacs}. 
To do so, for each calculation, 20 MD simulations were performed as described above,
each with a given value of the xenon-protein interaction parameter, $\lambda = \{ 0.0, 0.05 \cdots 0.95, 1.0 \}$, 
except that an harmonic restraint was added, namely, an intermolecular bond of type 6, between the xenon and a protein heavy atom, ten nsec of sampling being performed after the two first nsec of equilibration, the last five nsec being used for the actual free energy calculation.

In order to avoid singularities when the xenon atom is uncoupled ($\lambda = 0$), a soft-core potential was used \cite{Vangunsteren:94}, with a default radius of 0.3 nm,  the power in the soft-core potential being set to 1.0. 
In practice, the $\lambda$ interaction parameter was controlled through the vdw-lambdas and bonded-lambdas keywords. 

\section*{Results}

\subsection*{MD docking}

\begin{figure}[t!]
\includegraphics[width=8.0 cm]{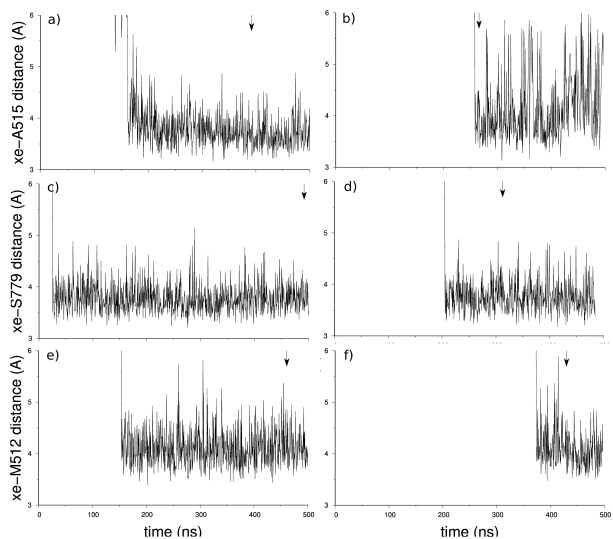}
\caption[]{
Distance between a xenon atom and an atom of the glycine binding domain of the NMDA receptor, as a function of time.
a) and b) Distance to the carbonyl oxygen of Ala 515 (site I);
c) and d) Distance to the $\gamma$-oxygen of Ser 779 (site II);
e) and f) Distance to the sulfur atom of Met 512 (site III);
a) Simulation A; b) and f) Simulation D; c) Simulation C; d) Simulation E; e) Simulation B.
The arrows indicate the time frames chosen for the calculations of the absolute binding free energy of xenon.	
}
\label{Fig:dxeoft}
\end{figure}
	
For each MD simulation, distances between each xenon atom and all protein heavy ones were monitored. As shown in Figure \ref{Fig:dxeoft}, six xenon atoms were found close (less than 6 {\AA} away) to a given protein atom for more than 100 nsec, and up to the end of the MD simulation.

\begin{figure}[t!]
\includegraphics[width=7.7 cm]{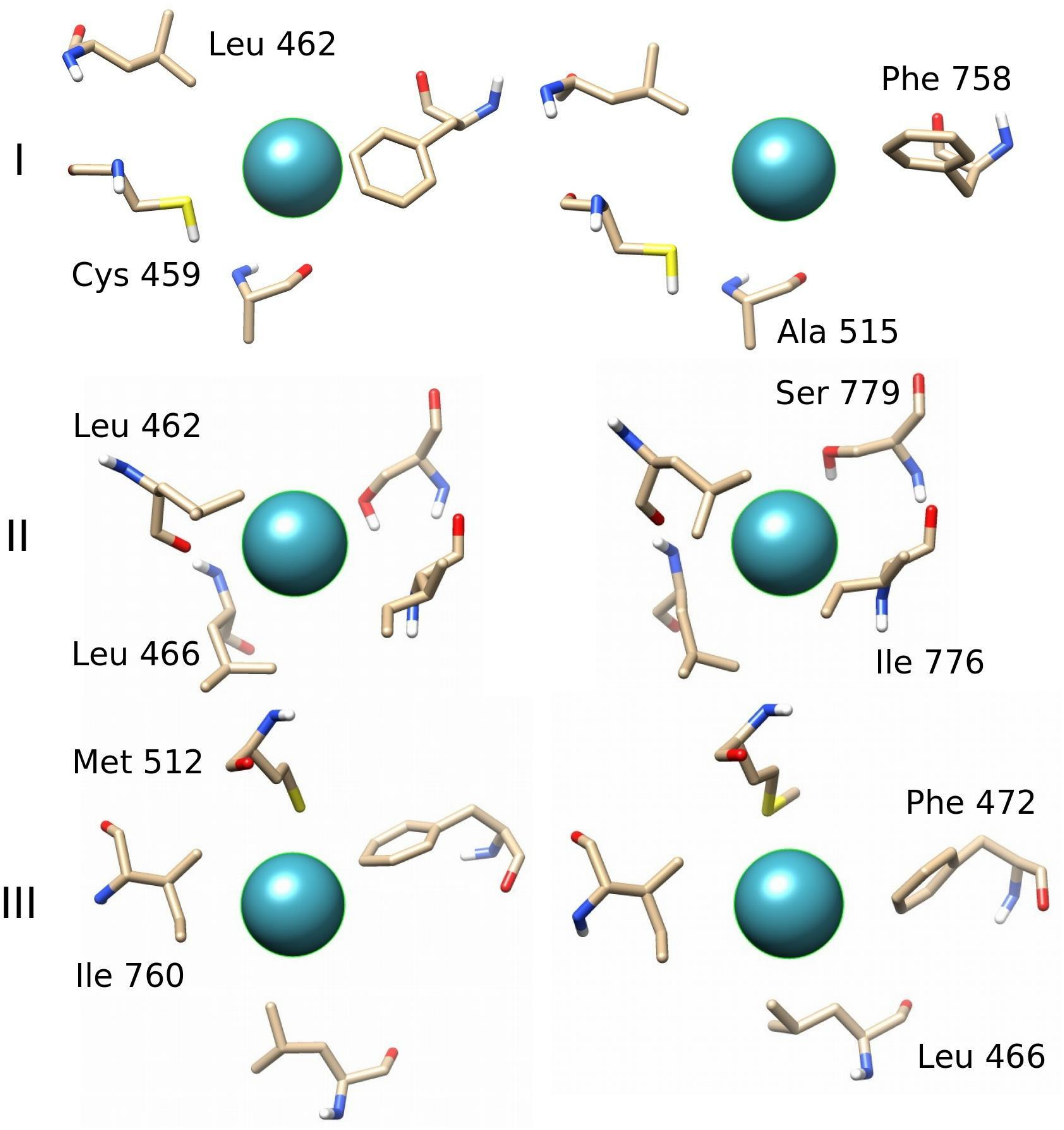}
\caption[]{
The three sites of xenon in the glycine binding domain of the NMDA receptor, as observed during MD simulations.
Top: site I (left: at t=392.5 ns of simulation A; right: at t=266.5 ns of simulation D); 
middle: site II (left: at t=492.5 ns of simulation C; right: at t=313.0 ns of simulation E); 
bottom: site III (left: at t=461.5 ns of simulation B; right: at t=430.0 ns of simulation D). The xenon atom is depicted as a sphere. Drawn with Chimera \cite{Chimera}. 	
}
\label{Fig:sitesxe}
\end{figure}

Interestingly, this corresponds to only three different binding sites (coined I to III), depicted in Figure \ref{Fig:sitesxe}, each of them being reached twice, two of them (sites I and III) being reached by two different xenon atoms during the same MD simulation (simulation D; Fig. \ref{Fig:dxeoft}b and \ref{Fig:dxeoft}f).    

Note that the sidechain of F758 is involved in site I. 
Note also that the three xenon binding sites are next to each other. 
Indeed, the sidechain of L462 is involved in both sites I and II,
while L466 is shared by sites II and III (see Fig. \ref{Fig:sitesxe}). 

On average, each xenon atom finds its binding site in $t_{fpt} \approx$ 200 ns (see Fig. \ref{Fig:dxeoft}), meaning that $k_{on}$, the binding constant, is of the order of \cite{Shaw:11c,Shaw:11d}: 
\[
k_{on} \approx \frac{1}{ t_{fpt} \textnormal{[Xe]}} = \textnormal{ 2 10$^8$ s$^{-1} \cdot$ M$^{-1}$}
\]
where $t_{fpt}$ is the mean first passage time, [Xe] being the xenon concentration during the MD simulations (25 mM$\cdot$L$^{-1}$).

\begin{table*}[t!]
\addtolength{\leftskip} {5mm}
 \caption{Absolute binding free energy of xenon (kJ$\cdot$mole$^{-1}$) in the three sites of the NMDA receptor found through MD docking. For each MD simulation, the time frame with the most representative xenon site configuration was chosen. During each thermodynamical cycle, the distance between the xenon and the closest protein heavy atom was restrained.}
 \label{Table:dG} 
\begin{tabular}{|c|c|c|c|c|c|c|c|}
 \hline
 Site & MD & Time (ns) & Restraint & Distance (nm) & $\Delta G^{prot}_{creation}$ & $\Delta G^{prot}_{unrestr}$ & $\Delta G^0_{binding}$ \\
 \hline
 \multirow{2}{*}{I}  & A & 392.5  & \multirow{2}{*}{A515:O} & \multirow{2}{*}{0.38} &  -5.5 $\pm$ 0.3 & -0.64 $\pm$ 0.04 & -2.6 $\pm$ 0.4 \\
 & D & 266.5 & & & -11.0 $\pm$ 0.5 & -0.36 $\pm$ 0.02 & -7.9 $\pm$ 0.5 \\
\hline
 \multirow{2}{*}{II} & C & 492.5 & \multirow{2}{*}{S779:O$_\gamma$}  & \multirow{2}{*}{0.38} & -17.0 $\pm$ 1.3 & -0.17 $\pm$ 0.01 & -13.7 $\pm$ 1.3 \\
 & E & 313.0 & & & -11.3 $\pm$ 0.6 & -0.16 $\pm$ 0.00 & -8.0 $\pm$ 0.6 \\
\hline
 \multirow{2}{*}{III} & B & 461.5 & \multirow{2}{*}{M512:S$_\delta$} & \multirow{2}{*}{0.41} & -15.5 $\pm$ 0.8 & -0.30 $\pm$ 0.01 & -12.3 $\pm$ 0.8 \\
 & D & 430.0 & & & -11.3 $\pm$ 0.5 & -0.30 $\pm$ 0.04 & -8.1 $\pm$ 0.5 \\
\hline
\end{tabular}
\end{table*}

\subsection*{ABFE calculations}

In order to assess the quality of the ABFE calculations analyzed hereafter, that is, both the quality of the force field used as well as the quality of the protocol considered (see Methods), $\Delta G^{solv}_{annihil}$, the free energy of transfer of the xenon atom from the bulk to the gas, was determined. 

The result obtained, namely: 
\[
\Delta G^{solv}_{annihil} = \textnormal{-4.3} \pm \textnormal{0.2 kJ}\cdot\textnormal{mole}^{-1}
\]
happens to be 20\% below the experimental value, namely, -5.4 kJ$\cdot$mole$^{-1}$ \cite{Marcus:84,Nau:18}. Such a small, though significant difference could be due to the choice of the water model, namely, TIP3P (see Methods). Indeed, results in fair agreement with experimental data have been obtained with the same xenon interaction parameters and either the SPC/E \cite{Masters:15} or the TIP4P/2005 models \cite{Vrabec:19}.  
However, the choice of the xenon parameters themselves, or of the mixing rules assumed for the description of xenon-water interactions \cite{Masters:15,Toennies:86}, may also play a role in this discrepancy.

For the ABFE calculations in the three putative xenon binding sites
obtained by MD docking, six representative NMDAR conformers were selected as follows.
First, for each MD simulation and each binding site, 
the protein heavy atom the closest to the xenon, on average, was identified, the
distance between both atoms being restrained during the ABFE calculation, so as to enforce the average value found during the corresponding MD simulation (see Table \ref{Table:dG}). Then, the three protein heavy atoms from other residues with the shortest distance fluctuations with respect to the xenon atom were identified. Next, the MD time frame for which the distances between the xenon and these four protein atoms are the closest to their average values was picked (indicated by arrows in Fig. \ref{Fig:dxeoft}).

As shown in Table \ref{Table:dG}, starting from this six NMDAR conformers,
the range of $\Delta G^0_{binding}$ values obtained is rather large, namely, between -2.6 $\pm$ 0.4 (site I, simulation A) and -13.7 $\pm$ 1.3 kJ$\cdot$mole$^{-1}$ (site II, simulation C). However, values around -8 kJ$\cdot$mole$^{-1}$ were obtained for each of the three binding sites (simulations D and E), suggesting that this significant variability is due to local conformational changes, probably in the vicinity of the xenon atoms. Indeed, as shown in Figure \ref{Fig:sitesxe}, several sidechains in the xenon binding sites happen to be in different rotameric states during the pair of ABFE calculations performed for a given site like, noteworthy, F758 (top of Fig. \ref{Fig:sitesxe}).
  
Note that such a dependence of the result of ABFE calculations upon the rotameric state of a sidechain in a binding site has already been observed \cite{Gapsys:21}. Note also that this means that larger $\Delta G^0_{binding}$ values could probably be obtained, if other NMDAR conformers were considered.

\subsection*{Conservation of the binding sites}

\begin{table}[t!]
 \caption{Conservation of the xenon binding sites in the glycine binding domain of the NMDA receptor.}
 \label{Table:cons} 
\begin{tabular}{|c|c|c|c|}
 \hline
 \multirow{2}{*}{Residue$^a$} & PDB            & \multirow{2}{*}{Site} & Conservation \\   
             & identifier$^b$ &      &   (\%) \\
 \hline
 Cys 459 &  67 & I      &  100.0 \\
 Leu 462 &  70 & I+II   &  100.0 \\
 Leu 466 &  74 & II+III &  100.0 \\
 Phe 472 &  80 & III    &  100.0 \\
 Met 512 & 120 & III    &   99.7 \\
 Ala 515 & 123 & I      &   99.7 \\
 Phe 758 & 250 & I+Gly  &   99.9 \\
 Ile 760 & 252 & III    &   99.9 \\
 Ile 776 & 268 & II     &   99.9 \\
 Ser 779 & 271 & II     &   97.2 \\
 \hline 
\end{tabular}
$^a$Uniprot rat (or human) sequence.\\
$^b$PDB 4KCC.\\
\vskip -0.5cm
\end{table}

The only presently available crystal structure of the apo form of the NMDAR glycine binding domain has been obtained with the rat sequence \cite{4kcc}.  
Note that, for this domain, the rat and human sequences are identical. More generally, as shown in Table \ref{Table:cons}, all residues shown in Figure \ref{Fig:sitesxe} are highly conserved, 
suggesting that the three xenon binding sites could prove involved in the function of the NMDAR.

\section*{Conclusion}

As a result of five independent MD docking simulations,
three different binding sites were found for xenon in the 
glycine binding site of the NMDAR.
Each site was found twice, that is, during 
two different simulations, two sites being found
during a single one (see Fig. \ref{Fig:dxeoft}).

All three xenon binding sites are next to each other,
each of them having at least a residue in common with
another one (see Fig. \ref{Fig:sitesxe}). 
Interestingly, site I is next to the glycine binding site,
the sidechain of F758 being involved in both binding sites.
This could explain why the F758W and F758Y mutations can prevent
competitive inhibition by xenon 
without affecting glycine binding \cite{Dickinson:12}.

For these sites, values of the absolute binding free energy of xenon 
up to -14 kJ$\cdot$mole$^{-1}$ were obtained (see Table \ref{Table:dG}).
However, since, for a given site, results were found to vary
significantly when the ABFE calculations
are performed for different protein conformers, larger values could probably be obtained, if other conformers of the glycine binding site of the NMADR were considered. 
 
On the other hand, since xenon could also compete
with glycine allosterically \cite{Jane:12,Paoletti:13}, 
the complete structure of the NMDAR needs also to be considered.
Work is in progress along these lines.
 

\end{document}